# Study and optimization of ion-irradiated High-Tc Josephson nanoJunctions by Monte Carlo simulations.

M. Sirena, N. Bergeal, J. Lesueur,
UPR5-CNRS, Physique Quantique, E.S.P.C.I., 10 Rue Vauquelin, 75231 Paris, France.

G. Faini
LPN-CNRS, Route de Nozay 91460, Marcoussis, France.

R. Bernard, J. Briatico, D. G. Crete and J.P. Contour.
UMR-CNRS/THALES, Route D128, 91767 Palaiseau, France.

Abstract.

High Tc Josephson nanoJunctions (HTc JnJ) made by irradiation have remarkable properties for technological applications. However, the spread in their electrical characteristics increases with the ion dose. We present a simple model to explain the JnJ inhomogeneities, which accounts quantitatively for experimental data. The spread in the slit's width of the irradiation mask is the limiting factor. Monte Carlo simulations have been performed using different irradiation conditions to study their influence on the spread of the JnJ characteristics. A "universal" behavior has been evidenced, which allows us to propose new strategies to optimize JnJ reproducibility.

Keywords: Josephson junctions, High Tc superconductor, Irradiation, Numerical simulations.

PACS: 74.78.Bz, 74.78.Na, 85.25Cp.

# I. INTRODUCTION.

In the last fifteen years there has been an intense research on High Tc Josephson Junctions (HTc JJ)[1]. The aims of these studies are related both to the comprehension of the basic physics (superconducting properties, electrical transport in superconducting heterostructures, etc …) and to technological developments for applications. The latter go from SQUIDs and Rapid Single Flux Quantum (RSFQ) devices to quantum voltage standards[2,3]. Many of these applications require closely packed series of hundreds of junctions with uniform $I_cR_n$ products, where $I_c$ is the critical current and $R_n$ the normal state resistance [4-6] of JJ.

Three techniques have been mainly used to make HTc JJ. One involves the deposition of superconductor/insulator heterostructures[7-9]. Their fabrication is complex, involve several lithography steps and suffers from pinholes in the thin insulating barrier[10,11] often shorting the superconducting electrodes. The other two techniques (leading to the so-called grain-boundary and edge junctions[12]) require either special substrates or delicate processing, which strongly limit their reproducibility and the successful operation of complex circuits. In the recent years, an alternative process has been proposed. Ion irradiation is used to disorder a superconducting layer, and therefore locally lower the critical temperature, named hereafter $T_c$'. Given the characteristic length scales in these materials, the irradiated part has to be on a nanometric scale for a sizable Josephson coupling to occur at temperatures above $T_c$'. Such a JJ is made of a microbridge of HTc material covered by a mask leaving a small aperture across (ranging from 20 to 100 nm depending on the experiments). High energy ions are used to create the local disorder through the slit : the dose sets Tc'. Above Tc' and up to a temperature called $T_J$, a Josephson coupling takes place between the two superconducting reservoirs, leading to a Superconductor/Normal/Superconductor (SNS) JJ.

The fabrication of HTc JJ by irradiation appears to be one of the most promising technique considering all the factors mentioned before. It allows the fabrication of closed packed arrays of several JJ at a nanoscale using a simple planar technology. The authors have achieved the fabrication of high quality Josephson nanoJunctions[12] (JnJ) with a very good reproducibility, thanks to the control and optimization of the fabrication process. However, it should be possible to further reduce the spread in the JJ properties. The aim of this work is to study and analyze the possible origins of this spread and to propose a route to reduce it.

**II. EXPERMENTAL DETAILS AND RESULTS.**

We have measured the transport properties of tens of HTc JnJ made in 150 nm thick c-axis oriented $YBa_2Cu_3O_{7-\delta}$ (YBCO) films grown on $SrTiO_3$ (STO) single crystals by Pulsed Laser Deposition (PLD). Details of the JnJ fabrication are given elsewhere[12]. The shadow mask used to define the junctions is made of polymethyl methacrylate (PMMA) photoresist ; 20 nm wide slits are opened by electron beam lithography. Figure 1 displays a picture of an array of JnJ together with a schematic view of the irradiated structure indicating the characteristic lengths and the x, y and z axis used along this work. The samples were irradiated with 100 keV oxygen ions with different doses ($\phi$) between 1.5 and 6 x $10^{13}$ ions/cm$^2$. The transport properties were measured by a standard four probe method between 4 K and 300 K.

Figure 2 presents the resistance (R) and the critical current (Ic) vs. temperature (T) for JnJ made on the same chip using different doses $\phi$: up to six different JnJ have been reported for each dose. The transition at high temperatures corresponds to the superconducting critical temperature $T_c$ of the pristine zones (reservoirs) and the lower transition corresponds to the Josephson coupling of the junctions themselves $T_J$. Figure 2 also shows that as the irradiation dose is increased $T_J$ decreases and the spread in $T_J$ increases as well. We do think that this latter effect is *intrinsically* related to the method used to fabricate the JnJ. The purpose of this work is to establish how the spread in JnJ characteristics is related to the irradiation process itself, and to propose a strategy to decrease it. It is worth noticing that the optimization of our current fabrication process has made our results so reproducible that this systematic behavior clearly appeared, and allowed us to make a quantitative study of the characteristics spread.

As we have said, the JJ operating temperature range is given by the Josephson regime ($T_c'<T<T_J$) and can be easily tuned by choosing the appropriate dose. The optimal temperature for a given application is driven by a set of factors, such as thermal noise, dependence of the critical current with temperature, cooling system, integration with other components in circuits and devices, etc ... Given the weak spread for JnJ irradiated at low dose, there is no need to improve it for $LN_2$ temperature operation. However, working temperatures between 30K and 50K show several advantages for RSFQ applications: the thermal noise is still low, and the system can be refrigerated with low power and low cost cryocoolers[6]. For this temperature range a reduction of the JJs dispersion is really an important issue.

What is required for applications is that both $I_c$ and $I_cR_n$ are the same for all the JJ at a given operating temperature $T_o$. As shown in Figure 2, the dispersion in $R_n$ from junction to junction is low, even for rather high dose. Moreover, the temperature dependence of $R_n$ is very weak. On the contrary, $I_c$ strongly varies with temperature (see Figure 2)[12], and is governed by the value of $T_J$. Therefore, a slight change in $T_J$ will produce a sizable change in $I_c$ for a given $T_o$. Since it has been shown that for the very same $T_J$, $I_c$ is reproducible from junction to junction[12], an accurate control of $T_J$ is the main requirement to decrease the spread in JnJ characteristics. It has been shown experimentally that for $T_J$ higher than 30K, $T_J$ is proportional to $T_c'$[13]. A model based on quasiclassical equations for diffusive inhomogeneous superconductors has been used to explain this effect[14]. For small enough variations, the dispersion in $T_J$ ($\Delta T_J$) can be taken to be proportional to that of $T_c'$ ($\Delta T_c'$), and therefore reducing $\Delta T_c'$ allows to meet our objective. Since YBCO has a d-wave superconducting order parameter, disorder decreases the critical temperature according to a depairing Abrikosov-Gorkov law[15]. It has been shown experimentally that the reduction of the transition temperature of the irradiated films ($T_c'$) with increasing irradiation damage (dpa stands for displacement per atom) in YBCO films can be expressed as[15]:

$$Ln\left(\frac{T_{C'}}{T_C}\right) = \psi\left(\frac{1}{2}\right) - \psi\left(\frac{1}{2} + \frac{0.14}{0.0375}\frac{dpa \cdot T_C}{T_{C'}}\right) \quad \text{(Eq. 1)}$$

where $\psi$ is the di-gamma function, and Tc the critical temperature of a virgin film. Based on this relationship, we have been able to compute the dispersion in $\Delta Tc'$, to make a quantitative comparison with the experimental data, and propose a new route towards lower spread in JJ characteristics.

We have used SRIM[16] Monte-Carlo simulations to model the geometry of the junction. This code provides the "lateral distribution of defects" LDD created by ions impinging a surface on one given spot: this is a local quantity, which depends on the coordinates (figure 1). In most of the experimental conditions used in this work (ion masses and energies), the thickness "e" of the sample is such that LDD does not vary a lot along the z-direction. We have therefore performed an integration of LDD along the z-direction, followed by integration on the width of the channel (W), to know C(y), the defect density in a slice $\Delta y$, along the y-direction corresponding to the extension of the junction. To know the defect density at a given y-coordinate, we have to add the contributions from ions impinging the whole length "d" of the junction which is physically the aperture width in the photoresist. Thus, dpa(y) is:

$$dpa(y) = \int_{-d/2}^{d/2} C(y - y_0) dy_0 \cdot \left(\frac{\phi}{\rho \cdot e}\right) = C^d(y) \cdot \frac{\phi}{\rho \cdot e} = C^d(y) \cdot \frac{dpa(0)}{C^d(0)} \quad \text{(Eq. 2)}$$

for an ion dose Φ and an atomic density ρ. $C^d(y)$ is the Integrated Lateral Distribution of Defects (ILDD). We can now express the dispersion in Tc' as a function of the parameters of the experiment.

In a simple and general way the variation of Tc' can be written as:

$$\Delta T_{C'} = \frac{\partial T_{C'}}{\partial dpa} \cdot \left( \frac{\partial dpa}{\partial d} \cdot \Delta d + \frac{\partial dpa}{\partial e} \cdot \Delta e + \frac{\partial dpa}{\partial W} \cdot \Delta W + \frac{\partial dpa}{\partial \phi} \cdot \Delta \phi \right) + \frac{\partial T_{C'}}{\partial T_C} \cdot \Delta Tc \quad \text{(Eq. 3)}$$

Basically ΔTc' is given by the dispersion of the damage done in different JJ (first term) and by the Tc variation from one JJ to an other. This second term has been written directly as a function of the $T_C$ gradient which includes all the sources of Tc dispersion (change in the sample thickness, chemical and structural in-homogeneities, etc ...) However, it is seen experimentally that the different Tc are the same within the experimental resolution (figure 2), indicating that ΔTc may be neglected. Considering that the JnJ are all irradiated at the same time with the same dose, Δϕ is negligible. Δe for the different JJ can also be considered to be small due to the close distance between the junctions (a few tens of microns, whereas several millimetres are required to see any sizable change in thickness for the samples grown by PLD). It can be easily seen that if the defect distribution width is much smaller than W the term $\partial dpa / \partial W$ is almost zero. In our case W is of the order of a few microns, more than one order of magnitude bigger than the typical LDD width (see for instance Figure 3). However, this is not the case for the JJ extension "d" which is of the order of 20 nm. Finally, and performing the derivative of equation 2, ΔTc' can be written as :

$$\Delta T_{C'} = \frac{\partial T_{C'}}{\partial dpa} \cdot \frac{C(d/2)}{C^d(0)} \cdot \Delta d \cdot dpa(0) \quad \text{(Eq. 4)}$$

The first factor can be called the "intrinsic" origin of the dispersion; it is due to the sensitivity of YBCO's critical temperature to irradiation damage (Eq. 1), which increases with dose, following the di-gamma function. For high doses a small variation of ϕ induces a great variation of Tc'. This factor is intrinsic to the system and there is nothing that can be done about it. Fortunately, the existence of the second factor allows to reduce ΔTc' (and therefore ΔT$_J$). The origin of this "extrinsic" factor is the dispersion in the slit's width (d) and it propagates to ΔTc' through the lateral damage distribution C(y). More precisely the proportionality factor is given by C(y) evaluated at the border of the slit (d/2) divided by the number of defects in the center of the slit ($C^d$(0)). A simple way to check this model is to consider a Gaussian lateral defect distribution of root mean square σ, and analyze the two extreme cases. When σ << d, C(d/2) ~ 0 and the model indicates there is almost no variation of T$_{C'}$ with d. On the other hand, if σ >> d, C(y) is practically constant between –d/2 and d/2 and dpa(0) is proportional to d. If we now modify d (d ± Δd), dpa(0) changes accordingly, and so does T$_{C'}$ and T$_J$. Under these conditions equation 4 becomes:

$$\Delta T_{C'} = \frac{\partial T_{C'}}{\partial dpa} \cdot \left[ dpa(0) \cdot \frac{\Delta d}{d} \right] \qquad (Ec.\ 5)$$

which correctly describes small changes in Tc' when C(y) is constant as a function of the lateral distance.

Summarizing, the dispersion in the slit's width (Δd) could be one of the main causes of the irradiated JnJ physical properties in-homogeneities, and the influence of this factor greatly depends on the parameter C(d/2). For this reason, a detailed study of the LDD becomes important, on the way of reducing the JnJ characteristic dispersion.

The defect's depth distribution in bulk materials and single films is well known[17,18], but little work has been done to study the LDD and its influence in the physical transport properties of irradiated JJ[19-22]. Recent developments in electron beam lithography to fabricate JJ at nanoscale require more accurate studies; we have used a numerical approach and Monte Carlo simulations to study the problem.

### III. NUMERICAL SIMULATIONS.

We have generated the lateral damage distribution using the Monte Carlo based code TRIM[16] for 150nm thick YBCO films over 1000 nm thick STO substrates. However, for very low irradiation energy the films thickness was reduced to keep an homogeneous distribution of defects within YBCO's depth. The defects density and position were calculated using full ion scattering cascades simulations, and sufficient statistics to get stationary results: typically around $10^6$ vacancies were created. Then C(y) was numerically integrated between $-d/2$ and $d/2$ to calculate dpa(y). We have chosen given irradiation conditions (ion mass, energy and dose), and calculated dpa(y) for different values of the JnJ extension ranging from d-Δd to d+Δd. Using equation 1, we have deduced ΔTc' for a given Δd.

The damage lateral distribution was simulated and ΔTc' calculated for He, O and Ar ions with irradiation energies ranging from 5 keV to 1MeV. More precisely, ΔTc' was calculated for O at different energies (30keV, 100keV, 200 keV, and 1000 keV), He (5 keV, 50 keV, and 100 keV) and for Ar ions (50, 100, 200 and 350 keV). We have computed the lateral damage distributions for slit size of d=19nm and d=21nm, and calculated ΔTc' as the difference between Tc'(19 nm) and Tc'(21 nm). This range in d was chosen because it reproduces quite well the measured ΔTc' for the different irradiation doses. Finaly, in an attempt to make even smaller JnJ and narrow lines and channels in HTSc, the dependence of the standard deviation (σ) of the defects distributions as a function of ion mass and energy was studied.

### IV. RESULTS.

Figure 3 shows the typical lateral damage distribution (C(y)) and the integrated damaged distribution ($C^d$(y)) made by O and He ion irradiation at 100 keV on a log-scale. The He lateral damage distributions are multiplied by ten for a better comparison with the O ones.

As expected, high energy light ions interact less with the superconductive layer, reducing the number of vacancies created per ion. This effect can be compensated by increasing the corresponding dose to finally have the same dpa and the same corresponding Tc'.

For both ions, the LDD presents two contributions : the first one rapidly falls off from the centre of the distribution while the other one appears as a round background, a "tail" that vanishes very slowly over long distances. For O ions, the damage density decreases from 22 to 6 defect/ion/nm in 7.5 nm but the second contribution keeps being important (1 defect/ion/nm) even 50 nm away from the center. For He ions, the second contribution is much lower and most of the vacancies are created at ± 20 nm from the centre. This "tail" effect is even more visible when integrating over the slit width to calculate the integrated lateral distribution. The Oxygen ILDD presents a rounded plateau due to the integration over d (20nm) that softens the first contribution, and the damage concentration decreases slowly without reaching saturation for distances as long as 100nm from the slit border (10 times the half-slit's size). On the other hand, He ions ILDD decreases very rapidly with increasing distance from the border of the slit, indicating that the distribution width is comparable to the slit size. The He ILDD presents a better defined shape controlled by the slit size. For the physical properties which are sensitive to the defects concentration such as the superconducting order parameter, this "tail" effect appears to be very important since the irradiation effects can be effective far from the slit border.

In general, the damage profile made by ion irradiation presents these two contributions but for low energy and heavy ions the second one is more important. The origin of this contribution is related to the damages done by backscattered ions. Olzierski et. al.[23] obtained similar results for electron irradiation in photoresists over different substrates. Their experiences and simulations displayed two contributions that can be fitted using two Gaussian distributions. These two contributions were ascribed to the "internal" proximity effect (forward scattered electrons and concentrated defects) and to the "external" proximity effect (back scattered electrons and extended defects distribution). Backscattered contribution in this system is even smaller than the one calculated for He, confirming the here observed dependence with the incident particle mass. A simple Gaussian distribution has been successful to get a more realistic description of the JnJ properties through a developed proximity effect model[14]. However including a more precise defect distribution can improve the quantitative results of the model. Moreover, a systematic study of the actual defect distribution form and width as a function of incident ion mass and energy needs to be done to optimize the junctions characteristics.

Using these realistic defects distributions, we have studied the validity of the simple model proposed in equation (4). Oxygen ions of 100 keV have been used to make JnJ with a nominal slit width of 20 nm ; for each dose we have measured Tc' and the corresponding $\Delta$Tc' as shown in Figure 2. Using the numerical simulations to calculate dpa(0) and equation (1) to evaluate $\frac{\partial T_{C'}}{\partial dpa}$, we have reported $\Delta T_{C'}$ as a function of the quantity $\frac{\partial T_{C'}}{\partial dpa} \cdot dpa(0)$ in Figure 4 (closed symbol). Within experimental error bars (15%), a clear linear relationship is observed, as expected from the model, indicating that the assumption of $\Delta$d-dominated fluctuations of $\Delta$Tc' is correct. Moreover, we have computed $\Delta T_{C'}$ following equation (4), by calculating the quantity $\delta = C(d/2)/C^d(0)$ for these irradiation conditions (open symbols). The only adjustable parameter to fit the experimental data is $\Delta$d, the slit width uncertaincy, which appears to be very low, on the order of 1 nm. This value seems a little bit small for standard electron beam lithography processes. Detailed experimental measurements in specially designed samples are on their way for a more precise comparison between the simulations and the experimental results[24]. If such an accuracy is confirmed, that may be a way to evaluate the precision of a given lithography process.

To go a step further in the evaluation of equation 4 validity, and with the goal of minimizing ΔTc', we have computed the quantity δ for a large set of ion masses and energies, and a slit of 20 nm. Then we have chosen a dose ϕ to target a Tc' for a given application ; doing so, $\partial T_C/\partial dpa$ ("intrinsic effect") and dpa(0) are also set. Using the experimental uncertainty Δd = 1nm, we have computed ΔTc', on the basis of the actual defect profile and equation (1). The result has been reported in Figure 5, where $\Delta T_{C'}$ has been plotted as a function of δ. The upper curve corresponds to $T_{C'}$ = 6.4K : it can be seen that the linear behavior expected from equation (4) is fully observed whatever the ion mass and energy are. Besides Tc', the only parameter which matters is δ, even in the extreme case reported here, where 1 MeV oxygen ions have been used ; for this last simulation, 1 μm thick film has been used in order to have an homogeneous damage profile in depth. Similar calculation have been performed for different values of Tc' and reported in Figure 5. Equation (4) always applies ; the slope of the curves decreases with decreasing dose, which means that the dispersion in $T_{C'}$ is less sensitive to δ, and therefore to irradiation conditions. In any case, reducing this factor minimizes $\Delta T_{C'}$. Coming back to the shape of the defects distributions, we can estimate the best situations to decrease it.

We have defined σ such as that 60% of the area under the LDD or the ILDD curve is within ±σ from its centre. Figure 6 shows σ as a function of the irradiation energy for He, O and Ar ions. In the three cases, σ presents a similar behavior. First, it increases as the energy increases, reaches a maximum value for E=$E_{Max}$ and then decreases as the energy increases. For low energies, all the damage is done in the superconducting film and, as the energy is increased the irradiation range increases. However, beyond a certain value (E>$E_{Max}$), the irradiation range becomes bigger than the film thickness and σ decreases. In this condition, and due to the irradiation profile (see Figure 6 inset), the damage done in the film concentrates along the irradiation axis. It is only in the substrate where the damage is extended proportionally to the energy. In this regime the defects become more concentrate as the energy is increased reducing the LDD width σ.

In the case of He and O ions, σ seems to saturate at very high energy. This range was not achieved for Ar ions in our simulation. Heavier ions need higher energies to obtain equivalent results. Although the behaviors observed here look similar for different ions, the curves do not scale with each other. In fact, for lighter ions the curves are more abrupt, indicating that the change with energy is more important. It is worth noting that decreasing σ for a given d, reduces δ since $C(d/2)$ is reduced more rapidly than $C^d(0)$. The optimization of the JJ can be ascribed in this way to the minimization of the LDD width (σ).

It is worthwhile noticing that σ values are bigger than the standard slit size for most of the cases (σ ≥ 15 nm) ; as a consequence using slits size smaller than 20 nm would not change much the ILDD width unless the LDD width is optimized before. We have calculated the ILDD widths for slits sizes of 20nm and 10nm and no significant change in σ was observed for most of the cases with the exception of O at 350 keV and He at 100 keV. For the last two cases the LDD width is smaller than the slit size (σ ≈ 8 nm) and the ILDD width is reduced by about 33% when the slit size is reduced from 20nm to 10nm. Small LDD widths can also be obtained using low energy irradiation ($E<E_{max}$) and the same considerations apply in this condition. However, in order to maintain a homogeneous damage distribution in depth, the film thickness must be reduced. Since it is well known[25] that superconducting properties can be substantially lowered when the thickness is reduced, there is a clear limitation in that direction. On the other hand, high energy ions (E > 150 keV for O and E > 50 keV for He) are very difficult to stop with standard photoresists. If high energy irradiation is needed it is necessary to change the lithography process in accordance to this, and for example to use multilayer structures to define the slit and stop the ions[9].

We see that it is possible to reduce $\Delta T_{C'}$ and to optimize the behavior of an array of several JnJ reducing the LDD width, more precisely reducing δ. As it was discussed before, there are two possibilities for this. The first one is to use high energy light ions, for which is needed to improve the stopping power of the mask used to define the slit. The other possibility is to use low energies; in this case a reduction in the film's thickness is needed in order to keep a homogeneous depth damage profile. Another possibility is to increase the slit's size d. As a matter of fact, increasing the slit size makes the ratio d/σ higher, reducing the dependence of the JJ's properties on Δd. This is clearly seen from the model; as d is increased, C(d/2) decreases and the ILDD between –d/2 and d/2 increases. Of course, this improves the JJ homogeneity but modifies the JJ properties. For example, for the same irradiating doses both dpa(0) and $T_{C'}$ will be increased. This can be easily corrected by adjusting ϕ. However, the Josephson coupling will be decreased, and the system will leave the true SNS behaviour, to enter the weak link limit. Experiments are on their way to study the progressive transition between this two limits and find the best optimized slit width. Such HTc JJ, with the irradiated part in the 100 nm range have been previously made[27]. The results are not close to what we are aiming for.

## V. CONCLUSIONS.

Ion irradiated Josephson NanoJunctions look very promising to develop new HTc superconductive electronics. The fabrication method we have set up gives results reproducible enough to quantitatively study the causes of the remaining spread in JJ characteristics.

We have developed a very simple and general model that explains the JJ in-homogeneities assuming that the main source of dispersion is the slit's size variation in the photoresist PMMA mask. The comparison with experimental data validates this assumption, and gives an upper limit for the uncertainty of the slit width made by electron beam lithography. The model indicates that $\Delta T_{C'}$ depends of two factors. The first is the "intrinsic" effect, which is given by the $T_{C'}$ dependence with the irradiation damage and a second, "extrinsic" effect which depends of the lateral damage distribution (LDD).

We have used Monte Carlo simulations (TRIM code) to generate the LDD and calculate $\Delta Tc'$ for different irradiation conditions in order to test the model. The distribution width as a function of energy presents a maximum related to the ions implantation range in the film and the substrate. Simulations results present a good agreement with the proposed "universal" model, presenting a linear dependence of $\Delta T_{C'}$ with δ. The slope of the curve is given by the "intrinsic" effect related to YBCO response to irradiation damage and the corresponding critical temperature $T_{C'}$. Data from different ions, energies and film thickness collapse on the same curve validating the generality of the model.

It is possible to reduce the JnJ in-homogeneities by reducing the LDD width. This can be done by two methods; reducing the ions energy and reducing accordingly the film thickness to maintain a homogeneous depth damage distribution or increasing the irradiation energy. This latter possibility appears to be the best one, but requires the development of more sophisticated masking techniques but may not be transferable to multilayers technologies.

**ACKNOWLEDGEMENT.**

The authors gratefully acknowledge O. Kaitasov for the ion irradiation made at IRMA-CSNSM (Orsay France), E Jacquet for the film growth and the personnel of the LPN-CNRS clean room for extraordinary technical support. M. Sirena acknowledges financial support from the MPPU- CNRS department.

**REFERENCES.**

CAPTIONS.

**Figure 1 :** Schematic view of a JnJ indicating the characteristic lengths in the x,y and z directions. The slit width "d" in the PMMA photoresist mask is the key factor. The top frame shows a picture of an array of 14 JnJ. The dark grey parts are superconducting microbridges ; the small black lines, the diffraction limited traces of the 20 nm slits ; the white areas are gold contact pads.

**Figure 2 :** Resistance (full lines) and critical current (dash line) vs. temperature for several JnJ irradiated with ϕ= 1.5, 3, 4.5 and 6 x $10^{13}$ ions/cm$^2$. Up to 6 JnJ per dose have been represented. For a given dose Φ, a coupling temperature $T_J$, and a dispersion $\Delta T_J$ can be defined.

**Figure 3:** Calculated Lateral Damage Distribution (LDD) (circles) and Integrated Lateral Damage Distribution (ILDD) (squares) for O and He of 100 keV. The ILDD have been normalized by the slits width and curves have been multiplied by different factors indicated in the figure for the sake of clarity. The lines are guides for the eyes.

**Figure 4 :** The spread in Tc' (ΔTc') vs. the quantity $\partial Tc'/\partial dpa \cdot dpa(0)$ (see text). Experimental data for 100 keV oxygen irradiation and a slit of nominal width 20 nm (closed symbol) align on a straight line. The proposed model (open symbol) quantitatively accounts for them if an uncertaincy of Δd = 1 nm on the slit width is used.

**Figure 5 :** Calculated $\Delta T_{C'}$ as function of δ (see text) for He, O and Ar irradiations at different energies, using Δd = 1 nm i. e. slit sizes d=21 nm and d=19 nm. The different lines correspond to linear fits according to the model proposed, for different Tc'. For Tc' = 56.9K and 74.1K, simulations were performed for an irradiation with oxygen ions only.

**Figure 6 :** Calculated lateral distribution width as a function of the irradiation energy for He, O, and Ar ions. The lines are a guide for the eyes. The insert shows a representation of the lateral damage distribution as a function of thickness for high energy irradiation (He at 100 keV), for a 100 nm thick YBCO film on STO.

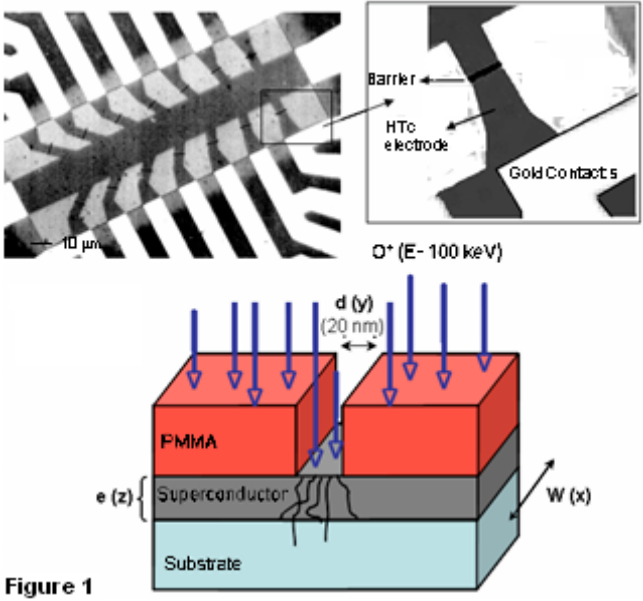

Figure 1

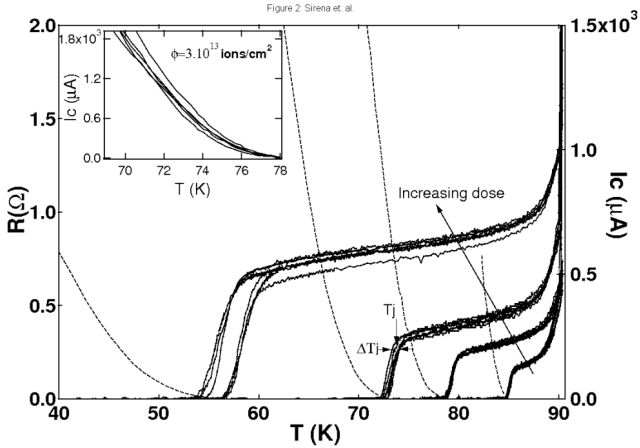

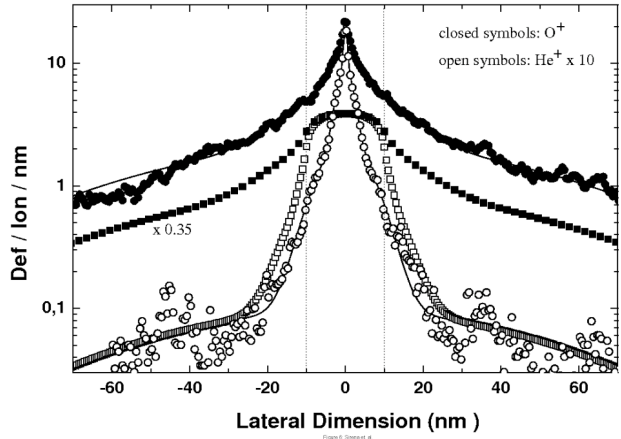

Figure 3: Sirena et. al.

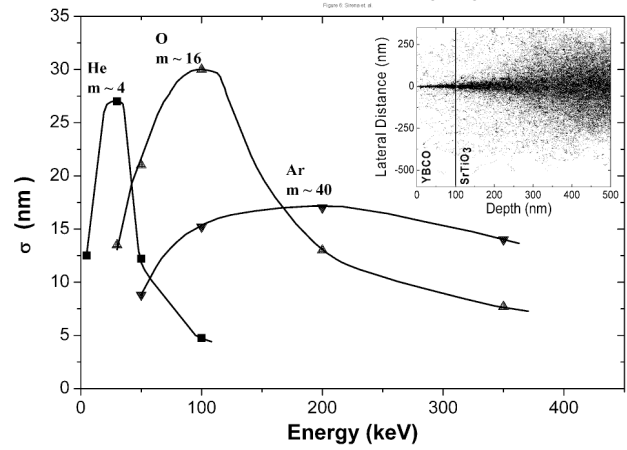

Figure 6: Sirena et. al.

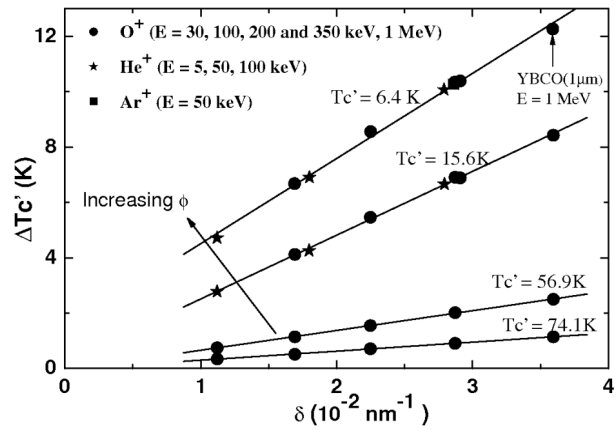

Figure 5: Sirena et. al.

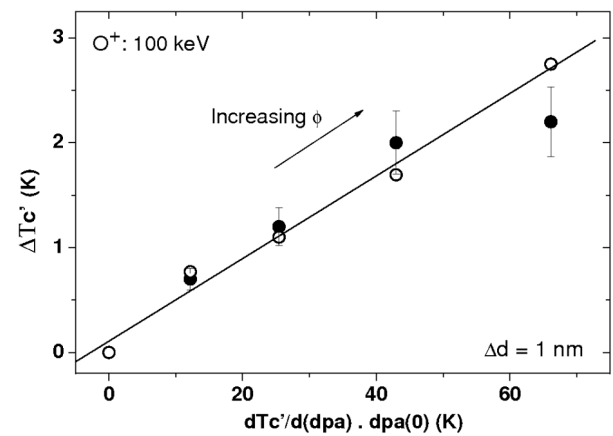

Figure 4: Sirena et. al.